\documentclass[twocolumn,showpacs,aps,prd,amsmath,amssymb]{revtex4-1}
\usepackage{graphicx}
\usepackage{xcolor}

\begin{document}

\title{%
 {\boldmath Observation of $\psi(3686) \to n\bar{n}$ and improved measurement of $\psi(3686) \to p \bar{p}$}}

%\author{Author list}
\author{
M.~Ablikim$^{1}$, M.~N.~Achasov$^{9,d}$, S.~Ahmed$^{14}$, M.~Albrecht$^{4}$, D.~J.~Ambrose$^{46}$, A.~Amoroso$^{51A,51C}$, F.~F.~An$^{1}$, Q.~An$^{48,39}$, J.~Z.~Bai$^{1}$, O.~Bakina$^{24}$
% ...
\\
\vspace{0.2cm}
(BESIII Collaboration)\\
\vspace{0.2cm} {\it
$^{1}$ Institute of High Energy Physics, Beijing 100049, People's Republic of China\\
% ...
$^{4}$ Bochum Ruhr-University, D-44780 Bochum, Germany\\
% ...
$^{9}$ G.I. Budker Institute of Nuclear Physics SB RAS (BINP), Novosibirsk 630090, Russia\\
% ...
$^{14}$ Helmholtz Institute Mainz, Johann-Joachim-Becher-Weg 45, D-55099 Mainz, Germany\\
% ...
$^{24}$ Joint Institute for Nuclear Research, 141980 Dubna, Moscow region, Russia\\
% ...
$^{39}$ State Key Laboratory of Particle Detection and Electronics, Beijing 100049, Hefei 230026, People's Republic of China\\
% ...
$^{46}$ University of Rochester, Rochester, New York 14627, USA\\
% ...
$^{48}$ University of Science and Technology of China, Hefei 230026, People's Republic of China\\
% ...
$^{51}$ (A)University of Turin, I-10125, Turin, Italy; (B)University of Eastern Piedmont, I-15121, Alessandria, Italy; (C)INFN, I-10125, Turin, Italy\\
% ...
$^{d}$ Also at the Novosibirsk State University, Novosibirsk, 630090, Russia.\\
% ...
}
}

\date{\today}

\begin{abstract}
  We observe the decay $\psi(3686) \to n \bar{n}$ for the first time and measure
  $\psi(3686) \to p \bar{p}$ with improved accuracy by using
  $1.07\times 10^8$ $\psi(3686)$ events collected with the BESIII detector. The
  measured branching fractions are $\mathcal{B}(\psi(3686) \to n
  \bar{n}) = (3.06 \pm 0.06 \pm 0.14)\times 10^{-4}$ and
  $\mathcal{B}(\psi(3686) \to p \bar{p}) = (3.05 \pm 0.02 \pm 0.12)
  \times 10^{-4}$.
  Here, the first uncertainties are statistical, and the second ones are
  systematic.
  With the hypothesis that the polar angular
  distributions of the neutron and proton in the center-of-mass system
  obey $1+\alpha \cos^2\theta$, we determine the $\alpha$ parameters
  to be $\alpha_{n\bar{n}} = 0.68 \pm 0.12 \pm 0.11$ and
  $\alpha_{p\bar{p}} = 1.03 \pm 0.06 \pm 0.03$ for $\psi(3686)\to
  n\bar{n}$ and $\psi(3686)\to p\bar{p}$, respectively.
\end{abstract}


\maketitle
\pagenumbering{arabic}
\vspace{10mm}

\section{Introduction}
As a theory of the strong interaction, QCD
has been well tested in the high energy region. However, in the lower
energy region, nonperturbative effects are dominant, and theoretical
calculations are very complicated.  The charmonium resonance
$\psi(3686)$ has a mass in the transition region between the perturbative and
nonperturbative regimes.  Therefore, studying $\psi(3686)$ hadronic
and electromagnetic decays will provide knowledge of its structure and
may shed light on perturbative and nonperturbative strong
interactions in this energy region~\cite{Asner:2008nq}. Nearly four
decades after the decay $\psi(3686) \to p \bar{p}$ was
measured~\cite{Feldman:1977nj}, we are able to measure $\psi(3686) \to
n \bar{n}$ for the first time using the large $\psi(3686)$ samples
collected at BESIII~\cite{Ablikim:2012pj}. A measurement of
$\psi(3686)\to n \bar{n}$, along with $\psi(3686) \to p \bar{p}$,
allows the testing of symmetries, such as flavor SU(3)~\cite{Zhu:2015bha}.

The measurements of $\psi(3686) \to N \bar{N}$, where $N$ represents a
neutron or proton throughout the text, allows the determination of the
relative phase angle between the amplitudes of the strong and
electromagnetic interactions.  The relative phase angle has been
studied via $J/\psi$ two-body decays to mesons with quantum numbers
$0^-0^-$~\cite{Suzuki:1999nb,LopezCastro:1994xw,Kopke:1988cs},
$1^-0^-$~\cite{Jousset:1988ni,Coffman:1988ve,LopezCastro:1994xw,Haber:1985cv},
$1^-1^-$~\cite{Kopke:1988cs,Adler:1987jy}, and $N
\bar{N}$~\cite{LopezCastro:1994xw,Baldini:1998en}. All results favor
near orthogonality between the two amplitudes. Recently, $J/\psi\to p
\bar{p}$ and $n \bar{n}$ have been measured by
BESIII~\cite{Ablikim:2012eu}, and confirm the previously measured
orthogonal relative phase angle. In contrast, experimental knowledge of
$\psi(3686)$ decays is relatively limited.  The decays of $J/\psi$ and
$\psi(3686)$ to same specific hadronic final states are naively
expected to be similar, and theoretical calculations favor a
relative phase of $90^\circ$ in $\psi(3686)$
decays~\cite{Gerard:1999uf}. However, the author of
Ref.~\cite{Suzuki:2001fs} argues that the relative phase angle in
decays to $1^-0^-$ and $1^+0^-$ final states is consistent with zero within the
experimental uncertainties for $\psi(3686)$ decays, and the
difference between $J/\psi$ and $\psi(3686)$ decays may be related to
a possible hadronic excess in $\psi(3686)$, which originates from a
long-distance process that is absent in $J/\psi$ decays. In contrast,
the authors of Refs.~\cite{Yuan:2003hj,Wang:2002np,Wang:2003hy}
suggest that the relative phase angle of $\psi(3686)$ decaying to
$1^-0^-$ and $0^-0^-$ final states could be large when the neglected
contribution from the continuum component is considered. Moreover, a
recent analysis based on previous measurements of $N\bar{N}$ final
states~\cite{Zhu:2015bha} suggests that there is a universal phase
angle for both $J/\psi$ and $\psi(3686)$ decays.  In short, no
conclusion can be drawn, and more experimental data are essential.

Also of interest for the processes of $e^+ e^- \to \psi(3686) \to N
\bar{N}$ is the angular distributions of the final states. The rate
of neutral vector resonance $V$ decaying into a particle-antiparticle
pair $h \bar{h}$ follows the distribution
$\mathrm{d}N/\mathrm{d}\cos\theta \propto 1+\alpha\cos^2\theta
$~\cite{Kessler:1970ef}, derived from the helicity formalism, where
$\theta$ is the polar angle of produced $h$ or $\bar{h}$ in the $V$
rest frame. Brodsky and Lepage~\cite{Brodsky:1981kj} predicted
$\alpha=1$, based on the QCD helicity conservation rule, which was
supported by an early measurement~\cite{Peruzzi:1977pb}.  However,
after a small $\alpha$ value for $J/\psi \to p \bar{p}$ was reported
with MARK II data (unpublished, mentioned in
Ref.~\cite{Claudson:1981fj}), later theoretical calculations, which
considered the effect of the hadron mass, suggested $\alpha$ might be
less than
$1$~\cite{Claudson:1981fj,Carimalo:1985mw,Murgia:1994dh,Bolz:1997as}.
Subsequent experiments supported this conclusion in $J/\psi$
decays~\cite{pdg2015}. For the decay of $\psi(3686) \to p \bar{p}$, as
shown in Table~\ref{tab:comp}, E835~\cite{Ambrogiani:2004uj} and
BESII~\cite{Ablikim:2006aw} have reported $\alpha$ values but with
large uncertainties, and both prefer to have an $\alpha$ less than
$1$. Up to now, there is no measurement of $\psi(3686) \to n
\bar{n}$. Besides the $N\bar{N}$ final states, $\alpha$ values have
been measured in other decay processes with baryon and antibaryon
pair final states, such as $J/\psi\to \Lambda\bar{\Lambda},
\Sigma\bar{\Sigma}^0$~\cite{Ablikim:2005cda}, $J/\psi\to
\Xi^+\bar{\Xi}^-, \Sigma(1385)\bar{\Sigma}(1385)$
~\cite{Ablikim:2012qn,Ablikim:2016sjb},
$\psi(3686)\to\Xi^+\bar{\Xi}^-,\Sigma(1385)\bar{\Sigma}(1385)$~\cite{Ablikim:2016iym,Ablikim:2016sjb},
and $J/\psi$ and $\psi(3686) \to \Xi^0
\bar{\Xi}^0$~\cite{Ablikim:2016sjb}. Unfortunately, no conclusive
theoretical model has been able to explain these measured $\alpha$
values.

\begin{table}[tbp]
\centering
\caption{\small Previous measurements of $\mathcal{B}(\psi(3686)\to p \bar{p})$ and
  $\alpha_{p\bar{p}}$.}
\label{tab:comp}
\begin{tabular}{ l  l  c }
\hline \hline
    & $\mathcal{B}$ (in $10^{-4}$) &  $\alpha$  \\
 \hline
%Our results & $3.05 \pm 0.02 \pm 0.13$ &  $1.03 \pm 0.06 \pm 0.03$ \\
World average~\cite{pdg2015}  & $2.88 \pm 0.10$  &  \\
World average (fit)~\cite{pdg2015}  &  $3.00 \pm 0.13 $ & \\
E835~\cite{Ambrogiani:2004uj} & & $0.67 \pm 0.15 \pm 0.04$ \\
BESII~\cite{Ablikim:2006aw} & $3.36\pm0.09\pm0.25$ & $0.85 \pm 0.24 \pm 0.04$ \\
CLEO~\cite{Pedlar:2005px} & $2.87\pm0.12\pm0.15$ & \\
BABAR~\cite{Lees:2013uta} & $3.14 \pm 0.28 \pm 0.18$ & \\
CLEOc data~\cite{Dobbs:2014ifa} & $3.08 \pm 0.05 \pm 0.18$ & \\
\hline \hline
\end{tabular}
\end{table}

Due to the Okubo-Zweig-Iizuka mechanism, the decays of $J/\psi$
and $\psi(3686)$ to hadrons are mediated via three gluons or a single
photon at the leading order. Perturbative QCD predicts the ``12\% rule,''
$Q_{h}=\frac{\mathcal{B}(\psi(3686)\to h)}{\mathcal{B}(J/\psi\to h)} =
\frac{\mathcal{B}(\psi(3686)\to\mu^+\mu^-)}{\mathcal{B}(J/\psi\to
  \mu^+\mu^-)} \approx 12.7\%$~\cite{12_rule_01,12_rule_02}. This rule
is expected to hold for both inclusive and exclusive processes but
was first observed to be violated in the decay of $\psi$ into
$\rho\pi$ by MARKII~\cite{Franklin:1983ve}, called the ``$\rho \pi$
puzzle.'' Reviews of the relevant theoretical and experimental
results~\cite{Gu:1999ks,Brambilla:2010cs,Wang:2012mf} conclude that the
current theoretical explanations are unsatisfactory. Further
precise measurements of $J/\psi$ and $\psi(3686)$ decay to $N\bar{N}$
may provide additional knowledge to help understand the $\rho\pi$
puzzle.

In this paper, we report the first measurement of $\psi(3686) \to n
\bar{n}$ and an improved measurement of $\psi(3686) \to p
\bar{p}$. First, we introduce the BESIII detector and the data samples used
in our analysis. Then, we describe the analysis and results of the
measurements of $\psi(3686) \to n \bar{n}$ and $\psi(3686) \to p
\bar{p}$. Finally, we compare the branching fractions and $\alpha$
values with previous experimental results and different theoretical
models.

\section{BESIII detector, data samples and simulation}
BESIII is a general purpose spectrometer with 93\% of $4\pi$ solid
angle geometrical acceptance~\cite{bes3}. A small cell, helium-based multilayer
drift chamber (MDC) provides momentum measurements of charged
particles with a resolution of 0.5\% at $1\ \mathrm{GeV}/c$ in a 1.0 T
magnetic field and energy loss ($d\mathrm{E}/dx$) measurements with a
resolution better than 6\% for electrons from Bhabha scattering. A
CsI(Tl) electromagnetic calorimeter (EMC) measures photon energies with a
resolution of 2.5\% (5\%) at $1\ \mathrm{GeV}$ in the barrel
(end caps). A time-of-flight system (TOF), composed of plastic
scintillators, with a time resolution of 80~ps (110~ps) in the barrel
(end caps) is used for particle identification (PID).  A
superconductive magnet provides a 1.0 T magnetic field in the central
region. A resistive plate chamber-based muon counter located in the
iron flux return of the magnet provides $2$~cm position resolution
and is used to identify muons with momentum greater than $0.5\
\mathrm{GeV}/c$. More details of the detector can be found
in Ref.~\cite{bes3}.

This analysis is based on a $\psi(3686)$ data sample corresponding to
$1.07\times 10^8$ events~\cite{Ablikim:2012pj} collected with the
BESIII detector operating at the BEPCII collider. An off-resonance data
sample with an integrated luminosity of $44\
\mathrm{pb}^{-1}$~\cite{Ablikim:2012pj}, taken at the c.m.\ energy of $3.65\ \mathrm{GeV}$, is used to determine the
non-$\psi(3686)$ backgrounds, {\it i.e.} those from nonresonant
processes, cosmic rays, and beam-related background.

A Monte Carlo (MC) simulated ``inclusive'' $\psi(3686)$ sample of
$1.07\times 10^8$ events is used to study the background.  The
$\psi(3686)$ resonance is produced by the event generator {\sc
  kkmc}~\cite{KKMC}, while the decays are generated by {\sc
  evtgen}~\cite{Lange:2001uf,EvtGen} for the known decays with the
branching fractions from the particle data group~\cite{pdg2015}, or by {\sc
  lundcharm}~\cite{Chen:2000tv} for the remaining unknown
decays. Signal MC samples for $\psi(3686)\to N\bar{N}$ are generated with
an angular distribution of $1+\alpha\cos^{2}\theta$, using the $\alpha$
values obtained from this analysis. The interaction of particles in
the detectors is simulated by a {\sc geant4}-based~\cite{geant4} MC
simulation software {\sc boost}~\cite{Deng:2006}, in which detector
resolutions and time-dependent beam-related backgrounds are
incorporated.

%%%%%%%%%%%%%%%%%%%%%%%%%%%%%%%%%%%%%%%%%%%%%%%%%%%%%%%%%%%%
%% n nbar part
%%%%%%%%%%%%%%%%%%%%%%%%%%%%%%%%%%%%%%%%%%%%%%%%%%%%%%%%%%%%

\section{Measurement of \boldmath $\psi(3686) \to n \bar{n}$}
The final state of the decay $\psi(3686) \to n \bar{n}$ consists of
a neutron and an
antineutron, which are back to back in the c.m.\ system and interact
with the EMC. The antineutron is expected to have higher interaction
probability and larger deposited energy in the EMC. To suppress
background efficiently and keep high efficiency for the signal, a {\sc
  root}-based~\cite{root}  multivariate analysis (MVA)~\cite{tmva}
is used.

\subsection{Event selection}
A signal candidate is required to have no charged tracks reconstructed
in the MDC. Events are selected using information from the EMC.
Showers must have deposited energy of $E>25\ \mathrm{MeV}$ in the
barrel ($|\cos\theta|<0.8$) or $E>50\ \mathrm{MeV}$ in the end caps
($0.86<|\cos\theta|<0.92$). The ``first shower'' is the most energetic
shower in the EMC, and the first shower group (SG) includes all
showers within a $0.9$ rad cone around the first shower. The
direction of a SG is taken as the energy-weighted average of the
directions of all showers within the SG. The SG's energy, number of
crystal hits and moments are the sums over all included showers for the relevant
variables. The ``second shower'' is the next most energetic shower
excluding the showers in the first SG, and the second SG is defined
based on the second shower analogous to how the first SG is
defined. The ``remaining showers'' are the rest of the showers which are
not included in the two leading SGs.

We require $|\cos\theta|<0.8$ for both SGs, and the energies of the
first SG and second SG to be larger than $600$ MeV and $60$ MeV,
respectively. The larger energy requirement applied to the first SG is
to select the antineutron, which is expected to have larger energy
deposits in the EMC than the neutron due to the annihilation of
the antineutron in the detector. There is a total of $6\times2
+2 = 14$ variables, which are listed in Table~\ref{tab:nnbar:mva:var},
including the energies, number of hits, second moments, lateral
moments, numbers of showers, largest opening angles of any two
showers within an SG, and number and summed energy of the
remaining showers.

\begin{table}[tbp]
\centering
\caption{\small The variables used in the MVA. The
  second moment is defined as $\sum_i E_i r^2_i / \sum_i E_i$, and the lateral moment is defined as
  $\sum_{i=3}^n E_i r^2_i / (E_1r^2_0 + E_2r^2_0 + \sum_{i=3}^n E_i r^2_i)$. Here,
  $r_0=5$ cm is the average distance between crystal centers in the EMC, $r_i$ is the radial
  distance of crystal $i$ from the cluster center, and $E_i$ is the crystal energy in
  decreasing order.}
\label{tab:nnbar:mva:var}
\begin{tabular}{l|l|c}
\hline \hline
Names  & Definitions & Importance \\
\hline
numhit1  & Number of hits in the first SG  & 0.09\\
numhit2  & Number of hits in the second SG & 0.06\\
ene1  & Energy of the first SG & 0.10\\
ene2  & Energy of the second SG & 0.21\\
secmom1  & Second moments of the first SG & 0.06\\
secmom2  & Second moments of the second SG & 0.06\\
latmom1  & Lateral moments of the first SG & 0.09\\
latmom2  & Lateral moments of the second SG & 0.05\\
bbang1 & Largest opening angle in the first SG & 0.04\\
bbang2 & Largest opening angle in the second SG & 0.05\\
numshow1 & Number of showers in the first SG & 0.04\\
numshow2 & Number of showers in the second SG & 0.04\\
numrem & Number of remaining showers  & 0.06 \\
enerem & Energy of remaining showers & 0.07 \\
\hline \hline
\end{tabular}
\end{table}

We implement the MVA by applying the boosted decision tree
(BDT)~\cite{bdt}. Here, $50\times 10^3$ signal and $100 \times 10^3$
background events are used as training samples. The signal events are
from signal MC simulation, and the background events are a weighted
mix of selected events from the off-resonance data at
$\sqrt{s}=3.65~\mathrm{GeV}$,
inclusive MC simulation, and exclusive MC simulation samples of
the processes $\psi(3686)\to \gamma \chi_{cJ}$, $\chi_{cJ} \to n \bar{n}$,
$(J=0,1,2)$, which are not included in the inclusive MC samples. The
scale factors are $3.7$ for the off-resonance data, determined based on
luminosity and cross sections~\cite{Ablikim:2012pj}, and $1.0$ for the
inclusive MC sample. We also select independent test samples
with the same components and number of events as the training samples.
The ``MVA'' selection criterion is obtained by the BDT method,
and it is optimized
under the assumption of $8900$ signal and $155,000$ background events,
which are estimated by a data sample within the
$\theta_{\mathrm{open}}>2.9$ radian region. Here
$\theta_{\mathrm{open}}$ is the opening angle between the two SGs in
the $e^+e^-$ c.m.\ system. Comparing training and testing samples, no
overtraining is found in the BDT analysis.  The chosen selection
criterion rejects approximately $95\%$ of the background while
retaining $76\%$ of all signal events.

\subsection{Background determination}
\label{sec:bkg:nnbar}

The signal will accumulate in the large $\theta_{\mathrm{open}}$
region since the final states are back to back. The possible peaking
background of $e^+e^-\to \gamma\gamma$ is studied with a MC sample of
$10^6$ events.
After the final selection, and scaled to the luminosity of real data,
only $27 \pm 10$ events are expected from this background source, which can be
neglected. This is also verified by studying the off-resonance
data. The remaining backgrounds are described by three components,
which are the same as those used in the BDT training.  None of
them produces a peak in the $\theta_{\mathrm{open}}$ distribution.

\subsection{Efficiency correction}
\label{sec:eff:cor}

\begin{figure*}[tbp]
\centering
\includegraphics[width=0.8\textwidth]{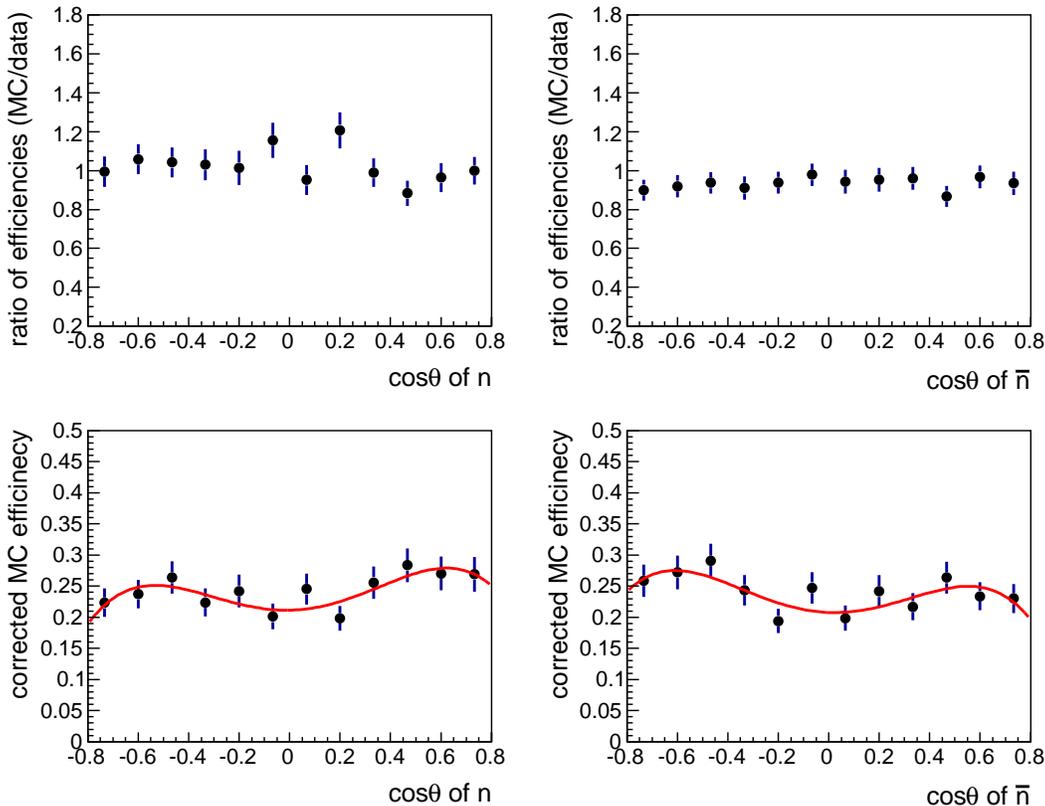}
\caption{(Top row) Ratios of the detection efficiencies between MC
  simulation and data vs $\cos\theta$ for neutron and
  antineutron, and
  (bottom row) the corrected detection efficiencies to select the
  $\psi(3686) \to n \bar{n}$ events vs $\cos\theta$. The solid
  curves are the fit results with a fourth-order polynomial
  function. The left plots are for the neutron, and the right ones are for the
  antineutron.}
\label{fig:eff:cor}
\end{figure*}

The neutron and antineutron efficiencies are corrected as a function
of $\cos\theta$ in the $e^+ e^-$ c.m.\ system to account for the
difference between data and MC simulation. Control samples of
$\psi(3686) \to p \bar{n} \pi^- + c.c.$, selected using charged tracks
only, are used to study this difference. The efficiency of the BDT
selector for the
antineutron is defined as $\epsilon = N_{\mathrm{BDT}} /
N_{\mathrm{tot}}$, where $N_{\mathrm{tot}}$ is the total number of
antineutron events obtained by a fit to the $p \pi$ recoil mass
distribution, and $N_{\mathrm{BDT}}$ is the number of antineutrons
selected with the BDT method. The same shower variables as used in
the nominal event selection are used in the BDT method to select the
antineutron candidate. The efficiency for the neutron is determined
analogously. The ratios of the efficiencies of MC simulation and data
as a function of $\cos\theta$ are assigned as the correction factors for the MC
efficiency of the neutron and antineutron, and are used to correct
the event selection efficiencies. The ratios and corrected
efficiencies are shown in Fig.~\ref{fig:eff:cor} for the neutron and
antineutron separately. The corrected efficiencies are fitted by
fourth-order polynomial functions with $\chi^2/ndf=0.87$ and $1.13$
for the neutron and antineutron, respectively.

\subsection{Branching fraction and angular distribution}
\label{sec:fit:nnbar}
\begin{figure}[tbp]
\centering
\includegraphics[width=0.45\textwidth]{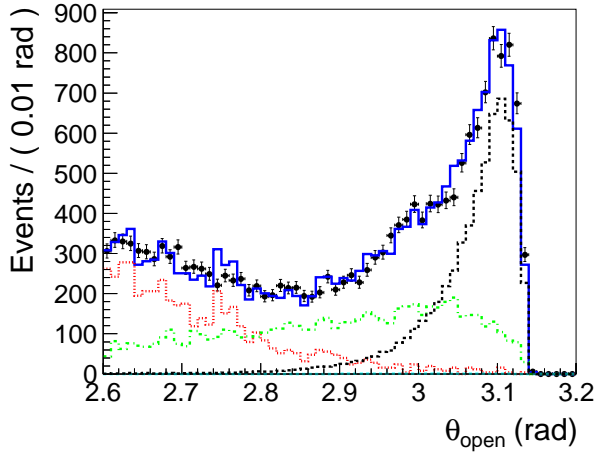}
\caption{Fit to the $\theta_{\mathrm{open}}$
  distribution. The data are shown by the dots with error bars. The fit
  result is shown as the solid blue curve. The signal shape is from MC
  simulation and is presented as the dashed black
  histogram. The background is described by three components:
  continuum background in dotted red, inclusive MC sample in
  dash-dotted green, and the tiny contribution from $\psi'\to\gamma
  \chi_{cJ}, \chi_{cJ}\to n \bar{n}$ (not included in the inclusive MC
  sample) in long-dashed cyan. All yields are free parameters in the
  fit.}
\label{fig:fit:nnbar:bbang}
\end{figure}

We perform a fit to the $\theta_{\mathrm{open}}$ distribution of data
to obtain the
numbers of signal candidates and background events.  The histogram
from signal MC simulation is used to construct the signal probability
density function (PDF). Corresponding histograms from the three background
components, as described in Sec.~\ref{sec:bkg:nnbar}, are used to
construct the background PDFs. The numbers of events from
each source are free parameters in the fit.
Figure~\ref{fig:fit:nnbar:bbang} shows the
fit to the $\theta_{\mathrm{open}}$ distribution. The fit yields $N_{sig}= 6056
\pm 117$ $n \bar{n}$ events with $\chi^2/ndf=3.24$.  Using a corrected
efficiency $\epsilon = 18.5\%$, the branching fraction of
$\psi(3686)\to n \bar{n}$ is determined to be $(3.06 \pm 0.06) \times 10^{-4}$ via $\mathcal{B} = N_{sig}/(N_{\psi(3686)}\epsilon)$,
where $N_{\psi(3686)}$ is the total number of $\psi(3686)$ and the uncertainty is statistical only.

\begin{figure*}[tbp]
\centering
\includegraphics[width=0.45\textwidth]{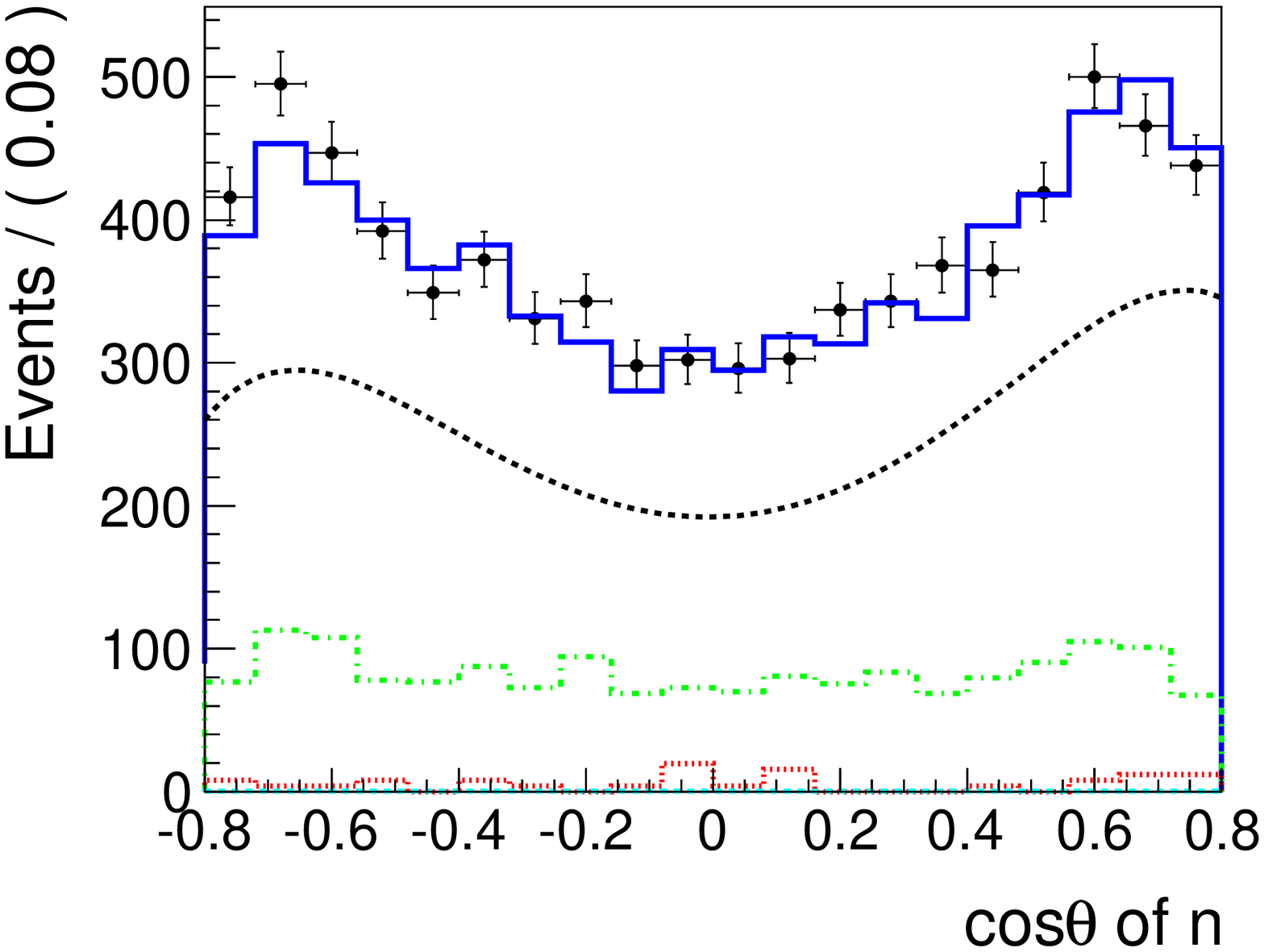}
\includegraphics[width=0.45\textwidth]{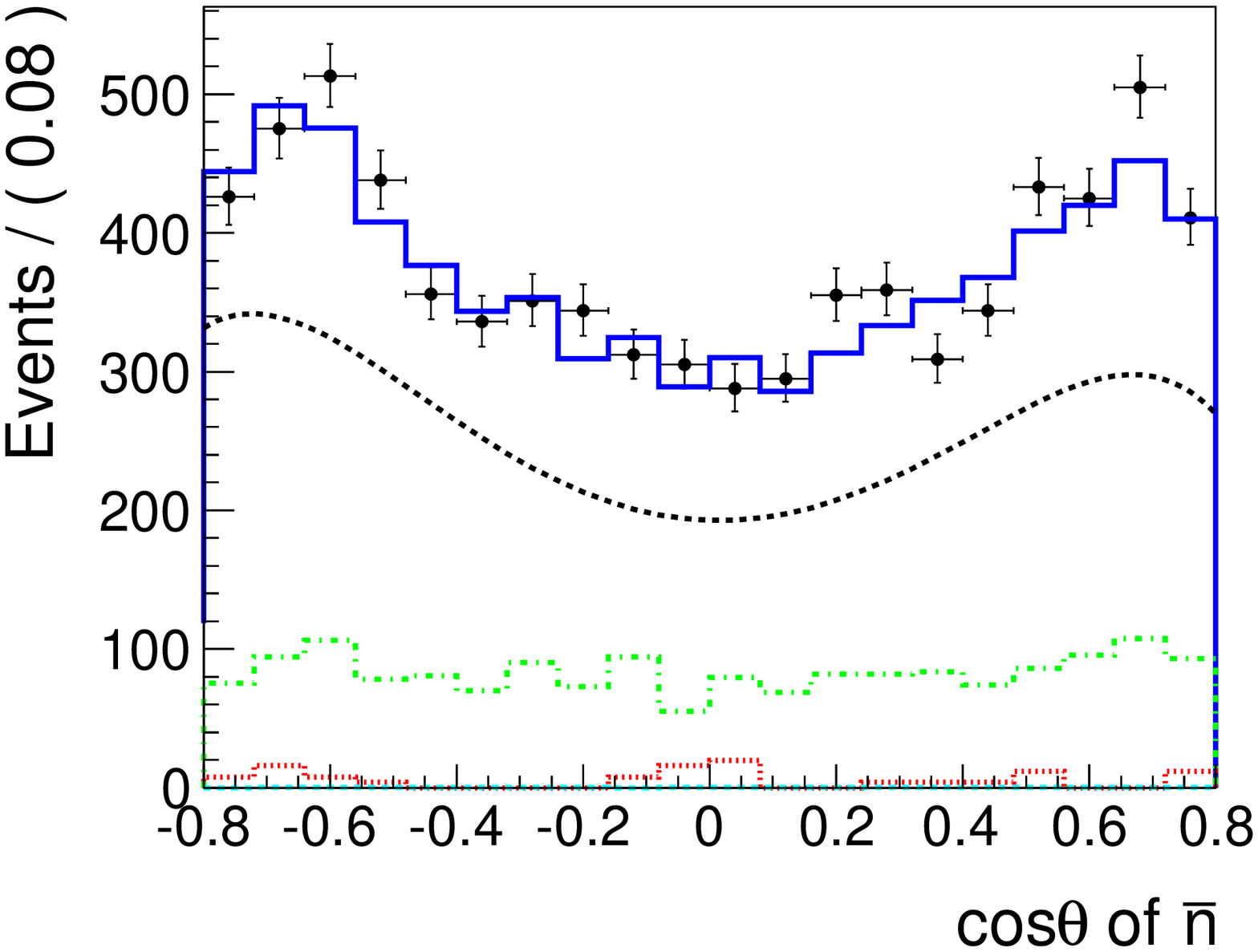}
\caption{Individual fits to the $\cos\theta$ distributions
  of (left) neutron and (right) antineutron. Data are shown as dots
  with error bars. The fit result is shown as the solid blue
  curve. The signal shape is parametrized by
  $(1+\alpha\cos^2\theta)\epsilon(\theta)$, shown as the
  dashed black curve. The background is described by three
  components: continuum background in dotted red, inclusive MC sample
  in dash-dotted green, and a small contribution from $\psi'\to\gamma
  \chi_{cJ}, \chi_{cJ}\to n \bar{n}$ in
  long-dashed cyan. }
\label{fig:fit:nnbar:alpha}
\end{figure*}

We fit the $\cos\theta_n$ and $\cos\theta_{\bar{n}}$ distributions
separately with fixed fractions of each component to determine the
$\alpha$ values. For these fits, an additional selection criterion
$\theta_{\mathrm{open}} >3.01$ is used to further suppress the continuum
background, and the fractions of each components within the region
$3.01 < \theta_{\mathrm{open}} < 3.20$ are obtained from the
$\theta_{\mathrm{open}}$ fit results. For the $\cos\theta_n$ and
$\cos\theta_{\bar{n}}$ distributions, the background PDFs are
constructed with the same method as used in the fits to
$\theta_{\mathrm{open}}$, while the signal PDF is constructed by the formula
$(1+\alpha\cos^2\theta)\epsilon(\theta)$. Here, $\epsilon(\theta)$ is
the corrected polar angle-dependent efficiency parameterized in a fourth-order polynomial, as described in
Sec.~\ref{sec:eff:cor}.  Figure~\ref{fig:fit:nnbar:alpha} shows the
fits to the $\cos\theta_n$ and $\cos\theta_{\bar{n}}$
distributions. An average $\alpha_{n\bar{n}}=0.68 \pm 0.12$ for the
angular distribution is obtained, while the separate fit results are
$0.76 \pm 0.12$ ($\chi^2/ndf=0.81$) and $0.60 \pm 0.12$
($\chi^2/ndf=2.01$) for the $\cos\theta_n$ and $\cos\theta_{\bar{n}}$
distributions, respectively. The uncertainties here are statistical
only. Since the neutron and antineutron are back to back in the c.m.\ %
system and the two angular distributions are fully correlated, the
average does not increase the statistics, and the uncertainty is not
changed.

\subsection{Systematic uncertainties}

\subsubsection{Resolution of $\theta_{\mathrm{open}}$}
To determine the difference in the $\theta_{\mathrm{open}}$ resolution
between data and MC, we fit the $\theta_{\mathrm{open}}$ distribution
of data with the signal PDF convolved with a Gaussian function of which the
parameters are left free in the fit. The
resultant mean and width of the Gaussian function are $0.005$ and
$0.002$ rad, respectively. With these modified PDFs, the resultant
changes are $0.3\%$ for the branching fraction and $0.0\%$ for the
$\alpha$ value, which are taken as the systematic uncertainties from
the resolution of $\theta_{\mathrm{open}}$. We do not consider the
resolution effect for the $\cos\theta$ distributions because of their
smoother shapes.

\subsubsection{Backgrounds}
The uncertainties associated with the background amplitudes are
estimated by fitting the $\theta_{\mathrm{open}}$ distribution with
fixed contributions for the continuum and inclusive MC background.
The differences between the new results and the nominal ones, $0.8 \%$ and
$8.1\%$ for the branching fraction and the $\alpha$ value, respectively, are
taken as the systematic uncertainties related with the background
amplitudes.

To estimate the effect on the $\alpha$ distribution from the continuum
background shape, we redo the fit to the $\cos\theta$ distributions
with the shape of the continuum background obtained without the
additional requirement $\theta_{\mathrm{open}}>3.01$, assuming that
there is no correlation between $\theta_{\mathrm{open}}$ and
$\cos\theta$. The difference of $\alpha$ to the nominal result is
$4.4\%$.

All in all, we determine the uncertainty from backgrounds to be $0.8\%$
for the branching fraction and $9.2\%$, the quadratic sum of $8.1\%$ and
$4.4\%$, for $\alpha$.

\subsubsection{Neutral reconstruction efficiencies}
 The reconstruction efficiency is corrected in bins of $\cos\theta$, and the
uncertainty of the correction is taken to be the statistical uncertainty,
which is about 2\% per
$\cos\theta$ bin. To obtain its effect on our results, we allow the efficiency to
fluctuate about the corrected efficiency according to the statistical
uncertainty, and redo the fits with the modified efficiencies.  We
also use the histograms of the corrected MC efficiencies directly. The
largest change of the signal yield is $0.2\%$ with the average
efficiency changing by $2\%$ ($1\%$ each from $\bar{n}$ and $n$), and
the largest change in $\alpha$ is $12.8\%$. We take these differences
from the standard results as the systematic uncertainties of the
neutral efficiency correction.

\subsubsection{Remaining showers}
We have checked and found that the number and energy of remaining showers are independent
of the angle, as we expected. Then only the branching fraction
measurement will be affected by the unperfect MC simulation.
Based on the distributions of the number and energy of remaining showers from
the data, we weight them in the signal MC considering their correlation. The difference is
found to be $0.4\%$ by comparing the efficiencies obtained with and without weighting, and is
quoted as the corresponding uncertainty.

\subsubsection{Analysis method}
We perform input/output checks by generating different signal MC
samples with different $\alpha$ values, from zero to unity; mixing
these signal MC samples with backgrounds; and scaling these samples to
the number of events according to data.  Compared to the input
values, the output signal yield is very close to the input, and
its corresponding systematic uncertainty can be neglected. For the
measurement of $\alpha$, the average difference, $2 \%$, is
taken as the systematic uncertainty.

\subsubsection{Binning}
In the nominal analysis, the $\theta_{\mathrm{open}}$, $\cos\theta_n$,
and $\cos\theta_{\bar{n}}$ distributions are divided into 60, 20 and
20 bins, respectively. To estimate the uncertainty associated with
binning, we redivide the distributions of $\theta_{\mathrm{open}}$,
$\cos\theta_n$, and $\cos\theta_{\bar{n}}$ into $[55,65]$, $[18,22]$,
$[18,22]$ bins, respectively and perform $11\times 5 \times 5 = 275$
fits of $\theta_{\mathrm{open}}$, $\cos\theta_n$, and
$\cos\theta_{\bar{n}}$ with all possible combinations of binnings to
determine the signal yields and $\alpha$ values. The differences between
the average results and the nominal values, $0.1\%$ for the branching
fraction and $4.5\%$ for the $\alpha$ value, are taken as the
systematic uncertainties.

\subsubsection{Physics model}
The signal efficiency in the branching fraction measurement depends
on the value of $\alpha$. Varying $\alpha$ by its standard
deviation, the relative change on the detection efficiency, $1.1\%$,
is taken as the systematic uncertainty due to the physics model.

\subsubsection{Trigger efficiency}
The neutral events used for this analysis are selected during data
taking by two trigger conditions:
1) the number of clusters in the EMC is required to be greater than
one,  and 2) the total energy deposited in the EMC is greater
than $0.5\ \mathrm{GeV}$~\cite{Berger:2010my}. The efficiency of the former
condition is very high~\cite{Berger:2010my}, and we conservatively
take $2\%$ as its systematic uncertainty.  Requiring the EMC total
energy to be larger than $0.9\ \mathrm{GeV}$, the trigger efficiency
of the second condition is $98.8\%$~\cite{Berger:2010my}, with an
uncertainty of $1.2\%$. Comparing the nominal results to the results
with the higher total energy requirement, the difference is
$0.2\%$. Combining the two gives $1.4\%$, which is taken as the
systematic uncertainty of the second trigger condition. Since these
two trigger conditions may be highly correlated, we take a
conservative $3.4\%$ as the total systematic uncertainty of the
trigger.

\subsubsection{Number of $\psi(3686)$ events}
The systematic uncertainty on the number of $\psi(3686)$ events is
$0.7\%$~\cite{Ablikim:2012pj}.

\subsubsection{Summary of systematic uncertainties}
The systematic uncertainties in the measurements of $\psi(3686) \to n
\bar{n}$ are summarized in Table~\ref{tab:nnbar:syserr}. Assuming
these systematic uncertainties are independent of each other, the
total uncertainty is obtained by adding the individual uncertainties
quadratically.

\begin{table}[tbp]
\centering
\caption{\small The relative systematic uncertainties for $\psi(3686) \to n \bar{n}$. Here "$\cdots$" denotes negligible.}
\label{tab:nnbar:syserr}
\begin{tabular}{l l l }
\hline \hline
Item                & Br ($\%$)        & $\alpha$ ($\%$)     \\
\hline
Resolution          & $0.3$   & $\cdots$            \\
Background          & $0.8$   & $9.2$      \\
Neutrals Efficiency       & $2.2$   & $12.8$      \\
Remaining Showers        & $0.4$   & $\cdots$           \\
Method              & $\cdots$         & $2.0 $     \\
Binning             & $0.1$   & $4.5$      \\
Physics model       & $1.1$   & $\cdots$    \\
Trigger             & $3.4$   & $\cdots$            \\
Number of $\psi'$      & $0.7$   & $\cdots$            \\
\hline
Total               & $4.4$   & $16.5$  \\
\hline \hline
\end{tabular}
\end{table}

%%%%%%%%%%%%%%%%%%%%%%%%%%%%%%%%%%%%%%%%%%%%%%%%%%%%%%%%%%%%
%% p pbar part
%%%%%%%%%%%%%%%%%%%%%%%%%%%%%%%%%%%%%%%%%%%%%%%%%%%%%%%%%%%%
\section{Measurement of \boldmath $\psi(3686) \to p \bar{p}$}
\subsection{Event selection}
The final state of $\psi(3686) \to p \bar{p}$ consists of a proton and an
antiproton, which are back to back and with a fixed momentum in the
c.m.\ system. A candidate charged track, reconstructed in the MDC, is
required to satisfy $V_r<1.0$ cm and $|V_z|<10.0$ cm, where $V_r$ and
$V_z$ are the distances of closest approach of the reconstructed track
to the interaction point, projected in a plane transverse to the beam
and along the beam direction, respectively. Two charged track
candidates with net charge zero are required. We also require the
momentum of each track to satisfy $1.546<p<1.628\ \mathrm{GeV}/c$ in
the c.m.\ system, which is within three times the resolution of the
expected momentum, and the polar angle to satisfy
$|\cos\theta|<0.8$. Using the information from the barrel TOF,
likelihoods $\mathcal{L}_i$ for different particle hypotheses are
calculated, and the
likelihood of both the proton and antiproton must satisfy
$\mathcal{L}_p>0.001$ and $\mathcal{L}_p>\mathcal{L}_K$,
where $\mathcal{L}_p$ is the PID likelihood for the
proton or antiproton hypothesis, and $\mathcal{L}_K$ is the
likelihood for the kaon
hypothesis. Further, we require the opening angle of the two tracks
to satisfy $\theta_{\mathrm{open}}>3.1\ \mathrm{rad}$ in the
$\psi(3686)$ c.m.\ system. There are $18,984$ candidate events satisfying
the selection criteria, which are used for the further study.

\subsection{Background estimation}
\begin{figure*}[tbp]
\centering
  \includegraphics[width=0.4\textwidth]{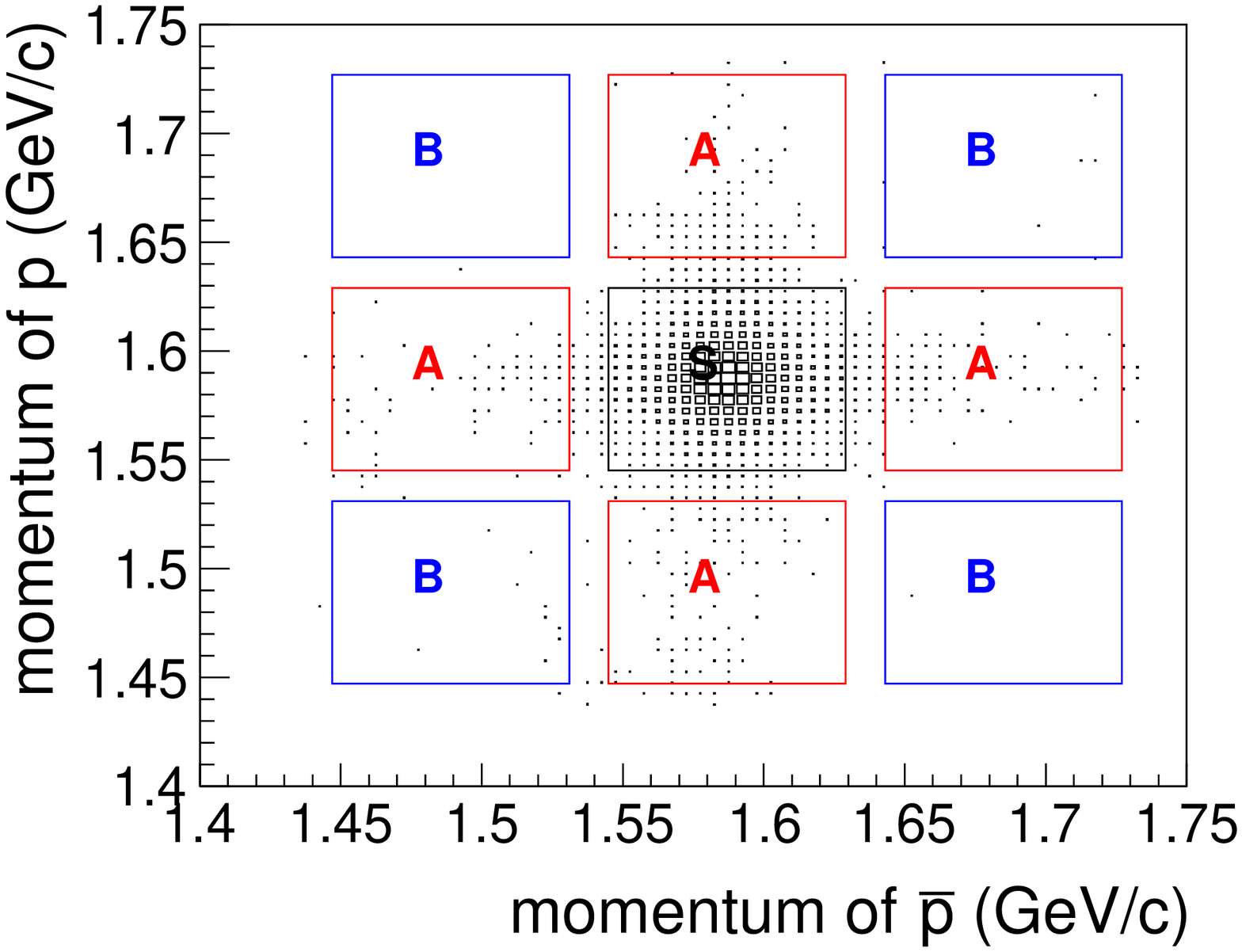}
  \includegraphics[width=0.4\textwidth]{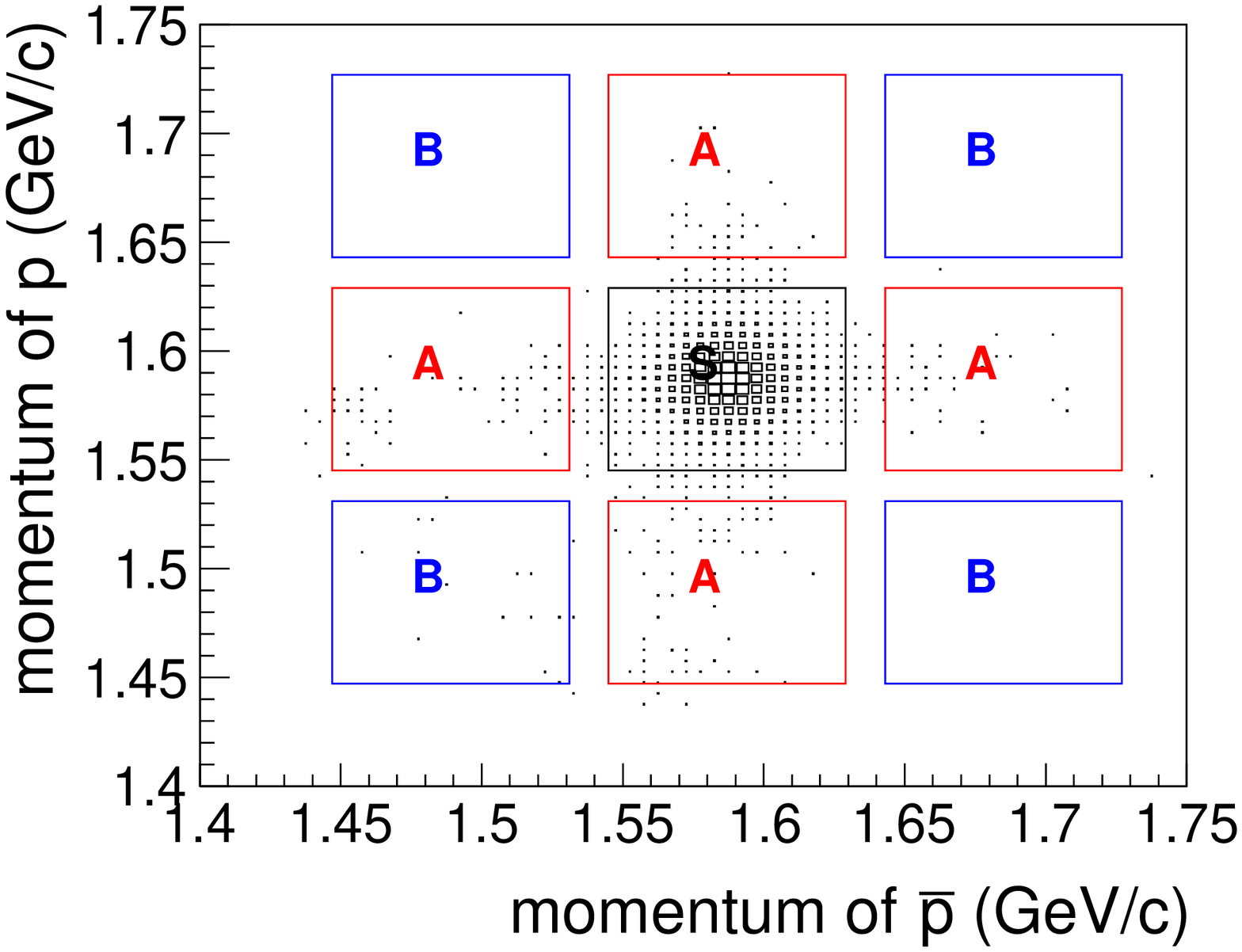}
  \caption{Scatter plots of momenta of proton vs antiproton. The
    left plot is data, and the right one is for inclusive MC.}
\label{fig:box}
\end{figure*}

In the analysis, backgrounds from the continuum process $e^+ e^- \to p
\bar{p}$ and $\psi(3686)$ decay into non-$p\bar{p}$ final states are
explored with different approaches. The former background is studied
with the off-resonance data at $\sqrt{s}=3.65$~GeV. With the same
selection criteria, there are $(22 \pm 5)$ events that survive, and
the expected background in the $\psi(3686)$ data is $(22 \pm 5) \times 3.7
= 81 \pm 18$ events, where $3.7$ is the scale factor which is the same
as in the $n\bar{n}$ study. By imposing the same selection criteria on
the $\psi(3686)$ inclusive MC sample, no non-$p\bar{p}$ final state
events survive, and the non-$p\bar{p}$ final state background from
$\psi(3686)$ decays is negligible. We also check the latter background
with the two-dimensional sidebands of the proton versus antiproton
momenta, which is shown in Fig.~\ref{fig:box}. There are a few events
in the sideband regions, marked as A and B in Fig.~\ref{fig:box}, but MC studies indicate that the events are
dominantly initial state or final state radiation events of
$\psi(3686) \to p \bar{p}$. The ratios of events in each sideband
region to that in signal region are consistent between data and signal
MC simulation.

\subsection{Efficiency correction}
\begin{figure*}[tbp]
\centering
  \includegraphics[width=0.8\textwidth]{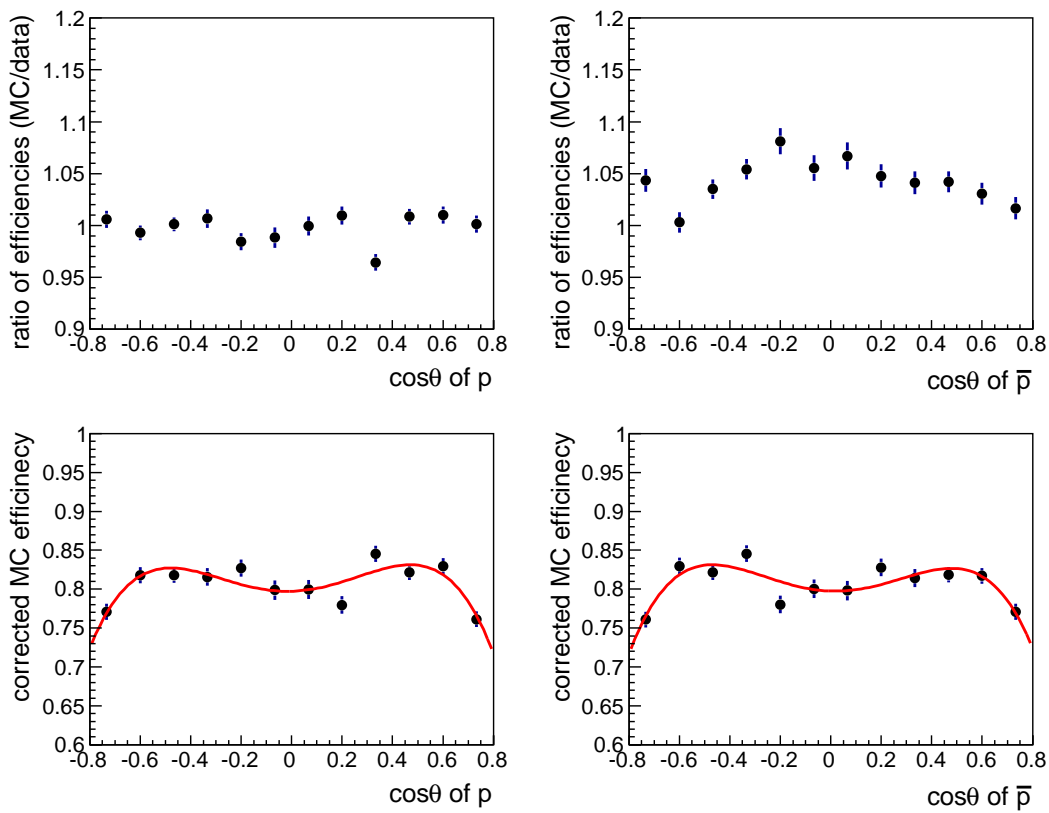}
  \caption{(Top row) Ratios of efficiencies of MC simulation over
    data and (bottom row) the corrected MC efficiency to select the
    signal events $\psi(3686) \to p \bar{p}$. The left plots are for
    the proton, and the right ones are for the antiproton.}
\label{fig:eff:ppbar}
\end{figure*}

In the $\psi(3686)\to p\bar{p}$ analysis, we correct the MC efficiency
as a function of $\cos\theta$ of the proton and antiproton, where the
corrected factors include both for tracking and PID
efficiencies. The efficiency differences between data and MC
simulation, which are obtained by studying the same control sample of
$\psi(3686) \to p \bar{p}$, are taken as the correction factors. To
determine the efficiency for the proton, we count the number of
$\psi(3686)\to p \bar{p}$ events by requiring an antiproton only, and
then check if the other track is reconstructed successfully in the
recoiling side and passes the PID selection criterion. The efficiency
is defined as $n_2/(n_1+n_2)$, where $n_1$ and $n_2$ are the yields of
events with only one reconstructed track identified as an antiproton
and with two reconstructed tracks identified as proton and
antiproton, respectively.  The yields $n_1$ and $n_2$ are obtained
from fits to the antiproton momentum distributions. In the fit, the
signal shape is described by the momentum distribution of the antiproton
with the standard selection criteria for $\psi(3686) \to p\bar{p}$,
and the background is described by a first-order polynomial function
since it is found to be flat from a study of the inclusive MC
sample. Cosmic rays and beam-related backgrounds are subtracted using
$V_z$-sidebands, in which $|V_z|\leq 5\ \mathrm{cm}$ is defined as the
signal region and $(-10<V_z<-5)$ and $(5<V_z<10)$ are defined as
sideband regions. A similar analysis is performed for the antiproton
detection efficiency. The ratio of efficiencies between MC simulation
and data are displayed individually in Fig.~\ref{fig:eff:ppbar} for the
proton and antiproton. We obtain the corrected MC efficiency to
select $\psi(3686) \to p \bar{p}$ candidates, also shown in
Fig.~\ref{fig:eff:ppbar}. The corrected MC efficiencies are
fitted with fourth-order polynomial functions with $\chi^2/ndf=2.56$
and $2.57$ for the proton and antiproton, respectively.

\subsection{Branching fraction and angular distribution}
\begin{figure*}[tbp]
\centering
  \includegraphics[width=0.4\textwidth]{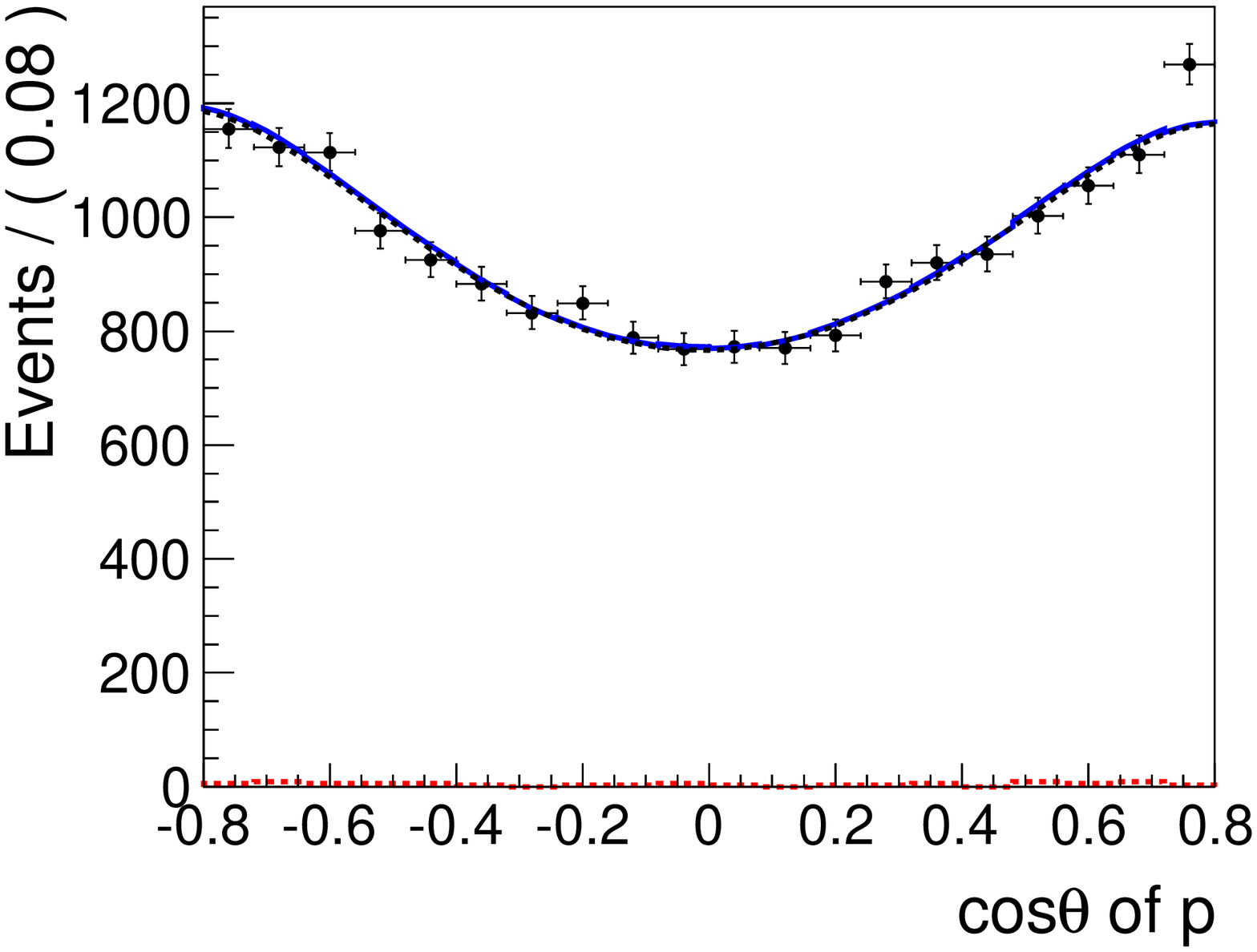}
  \includegraphics[width=0.4\textwidth]{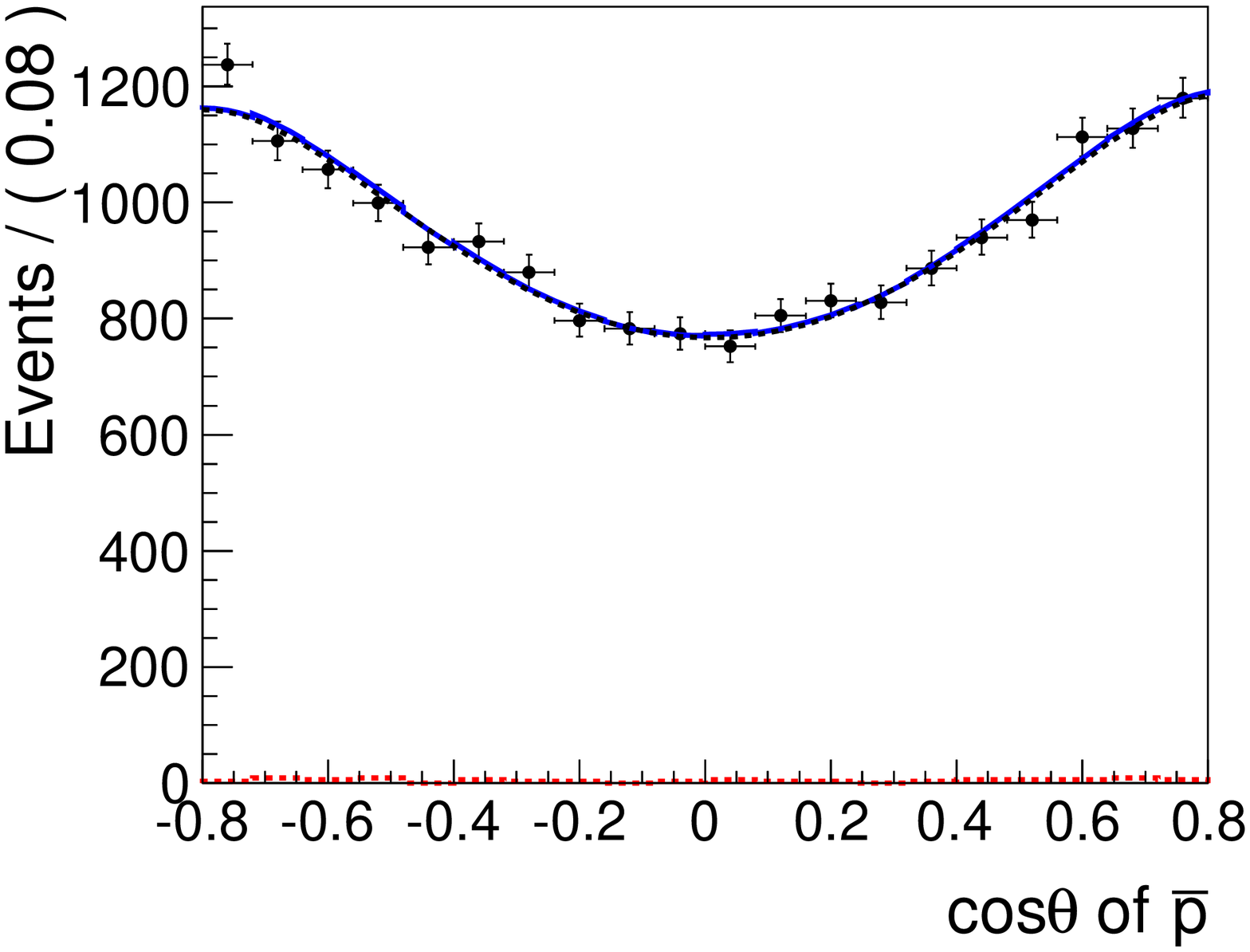}
  \caption{Fits to the $\cos\theta$ distributions of the (left) proton and
    (right) anti-proton. The dots with error bars are data, the solid
    blue lines are the fit curves, and the dashed red lines at the
    bottom of each plot are the backgrounds.}
\label{fig:measure:ppbar}
\end{figure*}

After subtracting the continuum background, the branching fraction is
determined to be $\mathcal{B}(\psi(3686)\to p \bar{p}) = (3.05 \pm
0.02) \times 10^{-4}$ via $\mathcal{B} = N_{sig}/(N_{\psi(3686)}\epsilon)$ with the corrected efficiency of $\epsilon =
58.1\%$ determined with the angular distribution corresponding to the
value of $\alpha$ obtained in this analysis. The $\cos\theta$
distributions of the proton and anti-proton for the selected candidates
are shown in Fig.~\ref{fig:measure:ppbar}. The distributions are
fitted with the functional form $N_\text{sig}(1+\alpha
\cos^2\theta)\epsilon(\theta) + N_\text{bg}f_\text{bg}$, where
$N_\text{bg}$ and $f_\text{bg}$, the yield and the shape of
the continuum background, are fixed in the fit according to the
off-resonance data at
$\sqrt{s}=3.65\ \mathrm{GeV}$. The fits are performed individually to the
$\cos\theta$ distributions of the proton and anti-proton and
yield the same value of $\alpha = 1.03 \pm 0.06$ with $\chi^2/ndf$
$1.06$ and $0.82$, respectively.

%%%%%%%%%%%%%%%%%%%%%%%%%%%%%%%%%%%%%%%%%%%%%%%%%%%%%%%%%%%%
%% systematic uncertainty of ppbar
%%%%%%%%%%%%%%%%%%%%%%%%%%%%%%%%%%%%%%%%%%%%%%%%%%%%%%%%%%%%

\subsection{Systematic uncertainties}
\subsubsection{Momentum resolution}
In this analysis, there are two requirements on the momentum,
$\theta_\text{open}>3.1$ and $1.546<p<1.628\ \mathrm{GeV}/c$, which
involve both its direction and magnitude.

We smear the momentum direction for the MC sample to improve the
consistency of the $\theta_\text{open}$ distributions between data and MC
simulation.  The detection efficiencies for the requirement
$\theta_{\mathrm{open}}>3.1$ are $98.1\%$ and $97.8\%$ without and
with the direction smearing, respectively. Thus, the systematic
uncertainty for the branching fraction measurement from this effect is
taken as $0.3\%$.

By fitting the momentum distributions of the proton and anti-proton,
the momentum resolutions are found to be $13.5$ and $11.2\
\mathrm{MeV}/c$ for data and MC simulation, respectively. The
corresponding efficiencies for the requirement $1.546<p<1.628\
\mathrm{GeV}/c$ are $99.76\%$ and $99.97\%$ for the data and MC
simulation, respectively, where the efficiencies are estimated by
integrating the Gaussian function within the specific signal
regions. Thus, the systematic uncertainty is taken to be $0.4\%$ for
the two charged tracks.

The total systematic uncertainty associated with the momentum
resolution for the branching fraction is $0.5\%$, and that for the
$\alpha$ value measurement is found to be negligible.

\subsubsection{Background}
The dominant background is from the continuum process, which is
estimated with the off-resonance data sample at $\sqrt{s}=3.65 \
\mathrm{GeV}$. The corresponding uncertainty of $18$ events, which is
$0.1\%$ of all signal events, is taken as the uncertainty in the branching
fraction measurement associated with the background. The uncertainty
on the $\alpha$ value associated with background is studied by
leaving the background yield free in the fit and found to be negligible.

\subsubsection{Tracking and PID efficiencies}
In the nominal analysis, the tracking and PID efficiencies for the
proton and anti-proton are corrected to improve the accuracy of the
measurement. Thus, only the uncertainties associated with the
statistics of correction factors and the method to exact correction
factors are considered.

We repeat the analysis $1000$ times by randomly fluctuating the
correction factors for the proton and anti-proton detection efficiency
with Gaussian functions independently in the different $\cos\theta$
bins, where the width of the Gaussian function is the statistical
uncertainty of the correction factors. The standard deviations of the
results are $<0.1\%$ for the branching fraction and $0.2\%$ for
$\alpha$, which are taken as the systematic uncertainties associated
with the statistical uncertainties.

In the nominal analysis, the corrected efficiency is parametrized
with a fourth-order polynomial function. Alternative parametrizations
with a polynomial function symmetric in $\cos\theta$ and directly
using the histogram for the corrected efficiency are performed. The
maximum changes of the branching fraction and $\alpha$ value, $3.3\%$
and $2.1\%$, respectively, are taken as the systematic uncertainties.

To be conservative, the linear sums of the two uncertainties, $3.3\%$
and $2.3\%$, are taken as the systematic uncertainties for the
branching fraction and $\alpha$ measurements associated with the
tracking and PID efficiency, respectively.

\subsubsection{Method}
From input/output checks, the average relative differences between
measured and true values are $1.1\%$ for the branching fraction and
$2.0\%$ for $\alpha$, which are taken as the systematic uncertainties.

\subsubsection{Binning}
In the nominal analysis, the $\cos\theta$ range of the proton and
anti-proton of $(-0.8,0.8)$ is divided into $20$ bins to determine the
corrected tracking and PID efficiency. Alternative analyses with $10$
or $40$ bins are also performed, and the largest differences with
respect to the nominal results are taken as the systematic
uncertainties associated with binning. The effect is negligible for
the branching fraction measurement and $1.0\%$ for the $\alpha$
measurement.

\subsubsection{Physics model}
In the branching fraction measurement, the detection efficiency
depends on the value of $\alpha$. Alternative detection efficiencies
varying $\alpha$ from $0.96$ to $1.10$, corresponding to
one standard deviation, are used. The largest change of the
efficiency with respect to the nominal value, $1.8\%$, is taken as the
systematic uncertainty.

\subsubsection{Trigger efficiency}
Events with two high momentum charged tracks in the barrel region of
the MDC have trigger efficiencies of $100.0\%$ and $99.94\%$ for
Bhabha and dimuon events~\cite{Berger:2010my}, respectively, and the
systematic uncertainty from the trigger is negligible.

\subsubsection{Number of $\psi(3686)$ events}
The systematic uncertainty on the number of $\psi(3686)$ events is
$0.7\%$~\cite{Ablikim:2012pj}.

\subsubsection{Summary of systematic uncertainties}
The systematic uncertainties of $\psi(3686) \to p\bar{p}$ from
the different sources  are summarized in
Table~\ref{tab:syserr:ppbar}. Assuming the systematic uncertainties
are independent, the total uncertainty is the sum on the individual
values added in quadrature.

\begin{table}[tbp]
  \centering
  \caption{\small Relative systematic uncertainties for the measurement of $\psi(3686)\to p \bar{p}$ in $\%$, where ``$\cdots$'' in the table means negligible.}
\label{tab:syserr:ppbar}
\begin{tabular}{ l  l  l }
\hline \hline
   & Br ($\%$) & $\alpha$  ($\%$)\\
 \hline
Resolution & $0.5$ & $\cdots$\\
Background & $0.1$ & $\cdots$\\
Tracking and PID & $3.3$ & $2.3$\\
Method & $1.1$ & $2.0$\\
Binning & $\cdots$ & $1.0$\\
Physics model & $1.8$ & $\cdots$\\
Trigger & $\cdots$ & $\cdots$\\
Number of $\psi(3686)$ & $0.7$ & $\cdots$\\
 \hline
Total & 4.0 & 3.2\\
\hline \hline
\end{tabular}
\end{table}

\section{Summary and Discussion}
In this paper, we measure the branching fractions of $\psi(3686) \to n
\bar{n}$ and $p \bar{p}$, and the $\alpha$ values of the polar angle
distribution, which are described by $1+\alpha\cos^2\theta$. The final
results are $\mathcal{B}(\psi(3686) \to n \bar{n}) = (3.06 \pm 0.06
\pm 0.14)\times 10^{-4}$ and $\alpha_{n\bar{n}} = 0.68 \pm 0.12 \pm
0.11$, and $\mathcal{B}(\psi(3686)\to p \bar{p}) = (3.05 \pm 0.02 \pm
0.12) \times 10^{-4}$ and $\alpha_{p\bar{p}} = 1.03 \pm 0.06 \pm
0.03$, where the former process is measured for the first time and
the latter one has improved precision compared to previous
measurements, as summarized in Table~\ref{tab:comp}. The measured
$\alpha_{p\bar{p}}$ is close to $1.0$, which is larger than previous
measurements, but both $\mathcal{B}(\psi(3686)\to p \bar{p})$ and
$\alpha_{p\bar{p}}$ are consistent with previous results within the
uncertainties.

To check for an odd $\cos\theta$ contribution from the $2\gamma$
exchange process~\cite{Pacetti:2015iqa}, we fit the angular
distributions as before but with the function $1 + \beta\cos\theta +
\alpha\cos^2\theta$. The results are $\beta_{n\bar{n}} = 0.04 \pm
0.05$ and $\beta_{p\bar{p}} = 0.01 \pm 0.02$. The possible
contributions from odd $\cos\theta$ terms in this analysis are
consistent with zero.

With the assumption the decay process is via a single photon exchange, the $\alpha$ value must satisfy $|\alpha|\le 1$~\cite{Faldt:2017kgy}. Then, the formula $1+\sin\phi\cos^2\theta$ is applied to fit to the $p\bar{p}$ data again, and we obtain the result $\phi_{p\bar{p}} = 1.57 \pm 0.28 \pm 0.25$,
where the statistical uncertainty is obtained from fit directly and the systematical uncertainty is propagated from the $3.2\%$ of the $\alpha_{p\bar{p}}$ value.

To compare with the 12\% rule, we use our measured branching
fractions to obtain
$$ \frac{\mathcal{B}(\psi(3686) \to p \bar{p})}{\mathcal{B}(J/\psi \to p \bar{p})} = (14.4
\pm 0.6)\% $$ and
$$ \frac{\mathcal{B}(\psi(3686) \to n \bar{n})}{\mathcal{B}(J/\psi \to n \bar{n})} = (14.8 \pm 1.2)\%,$$
where $\mathcal{B}(J/\psi \to p \bar{p})=(2.120 \pm 0.029)\times
10^{-3}$ and $\mathcal{B}(J/\psi \to n \bar{n}) = (2.09 \pm
0.16)\times 10^{-3}$ are the world average results~\cite{pdg2015}.
Both ratios are consistent with the 12\% rule.

In the decay of $J/\psi\to n\bar{n}$ and $p\bar{p}$~\cite{pdg2015},
both the branching fractions and $\alpha$ values are very close
between the two decay modes, which is expected if the strong
interaction is dominant in $J/\psi\to N\bar{N}$ decay and the
relative phase of between the strong and electromagnetic amplitudes is close to $90^\circ$
~\cite{Ablikim:2012eu}. In contrast, in $\psi(3686)$ decays, the
branching fractions are quite close between the two decay modes, but the
$\alpha$ values are not, which may imply a more complex mechanism in the
decay of $\psi(3686) \to N\bar{N}$. It makes a similar and straightforward extraction of the phase angle impossible in the decay of  $\psi(3686) \to N\bar{N}$, and further studies are deserved.

\acknowledgments
The BESIII Collaboration thanks the staff of BEPCII and the IHEP computing center for their strong support. This work is supported in part by National Key Basic Research Program of China under Contract No. 2015CB856700; National Natural Science Foundation of China (NSFC) under Contracts No. 11235011, No. 11335008, No. 11425524, No. 11625523, No. 11635010, and No. 11775246; the Ministry of Science and Technology under Contract No. 2015DFG02380; the Chinese Academy of Sciences (CAS) Large-Scale Scientific Facility Program; the CAS Center for Excellence in Particle Physics; Joint Large-Scale Scientific Facility Funds of the NSFC and CAS under Contracts No. U1332201, No. U1532257, No. U1532258, and No. U1632104; CAS under Contracts No. KJCX2-YW-N29, No. KJCX2-YW-N45, No. QYZDJ-SSW-SLH003; 100 Talents Program of CAS; National 1000 Talents Program of China; INPAC and Shanghai Key Laboratory for Particle Physics and Cosmology; German Research Foundation DFG under Contracts No. Collaborative Research Center CRC 1044 and No. FOR 2359; Istituto Nazionale di Fisica Nucleare, Italy; Koninklijke Nederlandse Akademie van Wetenschappen under Contract No. 530-4CDP03; Ministry of Development of Turkey under Contract No. DPT2006K-120470; National Natural Science Foundation of China under Contracts No. 11505034 and No. 11575077; National Science and Technology fund; The Swedish Research Council; U. S. Department of Energy under Contracts No. DE-FG02-05ER41374, No. DE-SC-0010118, No. DE-SC-0010504, and No. DE-SC-0012069; University of Groningen and the Helmholtzzentrum fuer Schwerionenforschung GmbH, Darmstadt; and WCU Program of National Research Foundation of Korea under Contract No. R32-2008-000-10155-010.


\begin{thebibliography}{99}
\bibitem{Asner:2008nq}
  D.~M.~Asner {\it et al.},
  %``Physics at BES-III,''
  Int.\ J.\ Mod.\ Phys.\ A {\bf 24}, S1 (2009)
  %%[arXiv:0809.1869 [hep-ex]].
  %%CITATION = ARXIV:0809.1869;%%
  %225 citations counted in INSPIRE as of 08 Dec 2016

\bibitem{Feldman:1977nj}
G.~J.~Feldman and M.~L.~Perl,
  %``Recent Results in electron-Positron Annihilation Above 2-GeV,''
Phys.\ Rep.\  {\bf 33}, 285 (1977).
  %%doi:10.1016/0370-1573(77)90024-2
  %%CITATION = doi:10.1016/0370-1573(77)90024-2;%%
  %171 citations counted in INSPIRE as of 08 Dec 2016

\bibitem{Ablikim:2012pj}
  M.~Ablikim {\it et al.} (BESIII Collaboration),
  %``Determination of the number of $\psi^{\prime}$ event at BESIII,''
  Chin.\ Phys.\ C {\bf 37}, 063001 (2013)
  %doi:10.1088/1674-1137/37/6/063001
  %%[arXiv:1209.6199 [hep-ex]].
  %%CITATION = doi:10.1088/1674-1137/37/6/063001;%%
  %37 citations counted in INSPIRE as of 10 Jul 2017

\bibitem{Zhu:2015bha}
  K.~Zhu, X.~H.~Mo, and C.~Z.~Yuan,
  %``Determination of the relative phase in $\psi'$ and $J/\psi$ decays into baryon and
  %antibaryon,''
  Int.\ J.\ Mod.\ Phys.\ A {\bf 30}, 1550148 (2015)
  %%[arXiv:1505.03930 [hep-ph]].
  %%CITATION = ARXIV:1505.03930;%%
  %1 citations counted in INSPIRE as of 06 Nov 2015

\bibitem{Suzuki:1999nb}
  M.~Suzuki,
  %``A large final state interaction in the 0- 0- decays of J / psi,''
  Phys.\ Rev.\ D {\bf 60}, 051501 (1999)
  %%[hep-ph/9901327].
  %%CITATION = HEP-PH/9901327;%%


\bibitem{LopezCastro:1994xw}
  G.~Lopez Castro, J.~L.~Lucio M. and J.~Pestieau,
  %``Tests of flavor symmetry in J / psi decays,''
  AIP Conf.\ Proc.\  {\bf 342}, 441 (1995)
  %%[hep-ph/9902300].
  %%CITATION = HEP-PH/9902300;%%

\bibitem{Kopke:1988cs}
  L.~K\"{o}pke and N.~Wermes,
  %``J/psi Decays,''
  Phys.\ Rep.\  {\bf 174}, 67 (1989).
  %%CITATION = PRPLC,174,67;%%

\bibitem{Jousset:1988ni}
  J.~Jousset {\it et al.}  (DM2 Collaboration),
  %``The J / Psi ---> Vector + Pseudoscalar Decays And The Eta, Eta-prime Quark Content,''
  Phys.\ Rev.\ D {\bf 41}, 1389 (1990).
  %%CITATION = PHRVA,D41,1389;%%


\bibitem{Coffman:1988ve}
  D.~Coffman {\it et al.}  (MARK-III Collaboration),
  %``MEASUREMENTS OF J / psi DECAYS INTO A VECTOR AND A PSEUDOSCALAR MESON,''
  Phys.\ Rev.\ D {\bf 38}, 2695 (1988); {\bf 40}, 3788(E) (1989).
  %%CITATION = PHRVA,D38,2695;%%


\bibitem{Haber:1985cv}
  H.~E.~Haber and J.~Perrier,
  %``A MODEL INDEPENDENT ANALYSIS OF HADRONIC DECAYS OF J / psi AND eta(c) (2980),''
  % NOTE: theory work of J/psi decays to PV
  Phys.\ Rev.\ D {\bf 32}, 2961 (1985).
  %%CITATION = PHRVA,D32,2961;%%

\bibitem{Adler:1987jy}
  J.~Adler {\it et al.}  (MARK-III Collaboration),
  % ``OBSERVATION OF   ELECTROMAGNETIC DECAYS OF J / psi TO VECTOR MESONS PAIRS'',
  in 1987 Europhys. Conference on High Energy Physics, Uppsala, Sweden, 1987 (unpublished).
  %%CITATION = INSPIRE-253449;%%

\bibitem{Baldini:1998en}
  R.~Baldini {\it et al.},
  %``Measurement of J / psi --> N anti-N branching ratios and estimate of the phase of the
  % strong decay amplitude,''
  Phys.\ Lett.\ B {\bf 444}, 111 (1998).
  %%CITATION = PHLTA,B444,111;%%

\bibitem{Ablikim:2012eu}
  M.~Ablikim {\it et al.}  (BESIII Collaboration),
  %``Study of $J/\psi\to p\bar{p}$ and $J/\psi\to n\bar{n}$,''
  Phys.\ Rev.\ D {\bf 86}, 032014 (2012)
  %%[arXiv:1205.1036 [hep-ex]].
  %%CITATION = ARXIV:1205.1036;%%
  %9 citations counted in INSPIRE as of 16 Dec 2013


\bibitem{Gerard:1999uf}
  J.~M.~Gerard and J.~Weyers,
  %``Phases and amplitudes in inclusive psi and psi-prime decays,''
  Phys.\ Lett.\ B {\bf 462}, 324 (1999)
  %%[hep-ph/9906357].
  %%CITATION = HEP-PH/9906357;%%


\bibitem{Suzuki:2001fs}
  M.~Suzuki,
  %``Possible hadronic excess in psi (2S) decay and the rho pi puzzle,''
  Phys.\ Rev.\ D {\bf 63}, 054021 (2001).
  %%CITATION = PHRVA,D63,054021;%%


\bibitem{Yuan:2003hj}
  C.~Z.~Yuan, P.~Wang and X.~H.~Mo,
  %``Relative phase between strong and electromagnetic amplitudes in psi(2S) ---> 0- 0-
  % decays,''
  Phys.\ Lett.\ B {\bf 567}, 73 (2003)
  %%[hep-ph/0305259].
  %%CITATION = HEP-PH/0305259;%%

\bibitem{Wang:2002np}
  P.~Wang, C.~Z.~Yuan, X.~H.~Mo and D.~H.~Zhang,
  %``The Interference between virtual photon and 1-- charmonium in e+ e- experiment,''
  Phys.\ Lett.\ B {\bf 593}, 89 (2004)
  %%[hep-ex/0210063].
  %%CITATION = HEP-EX/0210063;%%


\bibitem{Wang:2003hy}
  P.~Wang, C.~Z.~Yuan and X.~H.~Mo,
  %``Possible large phase in psi(2S) ---> 1- 0- decays,''
  Phys.\ Rev.\ D {\bf 69}, 057502 (2004)
  %%[hep-ph/0303144].
  %%CITATION = HEP-PH/0303144;%%


\bibitem{Kessler:1970ef}
  P.~Kessler,
  %``A generalized helicity method for feynman diagram calculations,''
  Nucl.\ Phys.\ {\bf B15}, 253 (1970).
  %%CITATION = NUPHA,B15,253;%%


\bibitem{Brodsky:1981kj}
  S.~J.~Brodsky and G.~P.~Lepage,
  %``Helicity Selection Rules and Tests of Gluon Spin in Exclusive QCD Processes,''
  Phys.\ Rev.\ D {\bf 24}, 2848 (1981).
  %%CITATION = PHRVA,D24,2848;%%

\bibitem{Peruzzi:1977pb}
  I.~Peruzzi {\it et al.},
  %``Baryonic Decays of the psi (3095),''
  Phys.\ Rev.\ D {\bf 17}, 2901 (1978).
  %%CITATION = PHRVA,D17,2901;%%
  %40 citations counted in INSPIRE as of 06 Dec 2013

\bibitem{Claudson:1981fj}
  M.~Claudson, S.~L.~Glashow and M.~B.~Wise,
  %``ISOSPIN VIOLATION IN J / psi ---> BARYON ANTI-BARYON,''
  Phys.\ Rev.\ D {\bf 25}, 1345 (1982).
  %%CITATION = PHRVA,D25,1345;%%

\bibitem{Carimalo:1985mw}
  C.~Carimalo,
  %``QUARK MASS EFFECTS IN psi ---> B anti-B DECAYS,''
  Int.\ J.\ Mod.\ Phys.\ A {\bf 02}, 249 (1987).
  %%CITATION = IMPAE,A2,249;%%


\bibitem{Murgia:1994dh}
  F.~Murgia and M.~Melis,
  %``Mass corrections in J / Psi ---> B anti-B decay and the role of distribution
  %  amplitudes,''
  Phys.\ Rev.\ D {\bf 51}, 3487 (1995)
  %%[hep-ph/9412205].
  %%CITATION = HEP-PH/9412205;%%

\bibitem{Bolz:1997as}
  J.~Bolz and P.~Kroll,
  %``Exclusive J / psi and psi-prime decays into baryon - anti-baryon pairs,''
  Eur.\ Phys.\ J.\ C {\bf 2}, 545 (1998)
  %%[hep-ph/9703252].
  %%CITATION = HEP-PH/9703252;%%


\bibitem{pdg2015}
% K.~A.~Olive et al.(Particle Data Group), Chin. Phys. C, 38, 090001 (2014) and 2015
% update.
C. Patrignani {\it et al.} (Particle Data Group), Chin. Phys. C, {\bf 40}, 100001 (2016) and
2017 update.

\bibitem{Ambrogiani:2004uj}
  M.~Ambrogiani {\it et al.}  (Fermilab E835 Collaboration),
  %``Measurement of the angular distribution in anti-p p ---> psi(2S) ---> e+ e-,''
  Phys.\ Lett.\ B {\bf 610}, 177 (2005)
  %%[hep-ex/0412007].
  %%CITATION = HEP-EX/0412007;%%

\bibitem{Ablikim:2006aw}
  M.~Ablikim {\it et al.}  (BES Collaboration),
  %``Measurements of psi(2S) decays to octet baryon-antibaryon pairs,''
  Phys.\ Lett.\ B {\bf 648}, 149 (2007)
  %%[hep-ex/0610079].
  %%CITATION = HEP-EX/0610079;%%

\bibitem{Pedlar:2005px}
  T.~K.~Pedlar {\it et al.}  (CLEO Collaboration),
  %``Branching fraction measurements of psi(2S) decay to baryon-antibaryon final states,''
  Phys.\ Rev.\ D {\bf 72}, 051108 (2005)
  %%[hep-ex/0505057].
  %%CITATION = HEP-EX/0505057;%%


\bibitem{Lees:2013uta}
  J.~P.~Lees {\it et al.}  (BaBar Collaboration),
  %``Measurement of the e+e- --> p anti-p cross section in the energy range from 3.0 to
  % 6.5 GeV,''
  Phys.\ Rev.\ D {\bf 88}, 072009 (2013)
  %%[arXiv:1308.1795 [hep-ex]].
  %%CITATION = ARXIV:1308.1795;%%
  %1 citations counted in INSPIRE as of 11 Dec 2013

\bibitem{Dobbs:2014ifa}
  S.~Dobbs, A.~Tomaradze, T.~Xiao, K.~K.~Seth and G.~Bonvicini,
  %``First measurements of timelike form factors of the hyperons, and evidence of diquark correlations,''
  Phys.\ Lett.\ B {\bf 739}, 90 (2014)
  % doi:10.1016/j.physletb.2014.10.025
  %%[arXiv:1410.8356 [hep-ex]].
  %%CITATION = doi:10.1016/j.physletb.2014.10.025;%%
  %1 citations counted in INSPIRE as of 05 Jan 2016


\bibitem{Ablikim:2005cda}
  M.~Ablikim {\it et al.} (BES Collaboration),
  %``Study of J / psi decays to Lambda anti-Lambda and Sigma0 anti-Sigma0,''
  Phys.\ Lett.\ B {\bf 632}, 181 (2006)
  %%doi:10.1016/j.physletb.2005.10.079
  %%[hep-ex/0506020].
  %%CITATION = doi:10.1016/j.physletb.2005.10.079;%%
  %19 citations counted in INSPIRE as of 22 Dec 2016


\bibitem{Ablikim:2012qn}
  M.~Ablikim {\it et al.} (BESII Collaboration),
  %``Measurement of J/psi decays into Lambda anti-Lambda pi+ pi-,''
  Chin.\ Phys.\ C {\bf 36}, 1031 (2012).
  %%doi:10.1088/1674-1137/36/11/001
  %%CITATION = doi:10.1088/1674-1137/36/11/001;%%
  %2 citations counted in INSPIRE as of 22 Dec 2016


\bibitem{Ablikim:2016sjb}
  M.~Ablikim {\it et al.} (BESIII Collaboration),
    %``Study of $J/\psi$ and $\psi(3686)\rightarrow\Sigma(1385)^{0}\bar\Sigma(1385)^{0}$
  %and $\Xi^0\bar\Xi^{0}$,''
  Phys.\ Lett.\ B {\bf 770}, 217 (2017)
  %%doi:10.1016/j.physletb.2017.04.048
  %%[arXiv:1612.08664 [hep-ex]].
  %%CITATION = doi:10.1016/j.physletb.2017.04.048;%%
  %1 citations counted in INSPIRE as of 27 Jul 2017


\bibitem{Ablikim:2016iym}
  M.~Ablikim {\it et al.} (BESIII Collaboration),
  %``Study of $\psi$ decays to the $\Xi^{-}\bar\Xi^{+}$ and $\Sigma(1385)^{\mp}\bar\Sigma(1385)^{\pm}$ final states,''
  Phys.\ Rev.\ D {\bf 93}, 072003 (2016)
  %%doi:10.1103/PhysRevD.93.072003
  %%[arXiv:1602.06754 [hep-ex]].
  %%CITATION = doi:10.1103/PhysRevD.93.072003;%%
  %1 citations counted in INSPIRE as of 22 Dec 2016


\bibitem{12_rule_01} T. Appelquist and H. D. Politzer, Phys. Rev. Lett., $\bf{34}$, 43 (1975).
\bibitem{12_rule_02} A. De R\`ujula and S. L. Glashow, Phys. Rev. Lett., $\bf{34}$, 46 (1975).

\bibitem{Franklin:1983ve}
  M.~E.~B.~Franklin {\it et al.},
    %``Measurement of $\psi(3097)$ and $\psi^\prime$ (3686) Decays Into Selected Hadronic
  %Modes,''
  Phys.\ Rev.\ Lett.\  {\bf 51}, 963 (1983).
  %%doi:10.1103/PhysRevLett.51.963
  %%CITATION = doi:10.1103/PhysRevLett.51.963;%%
  %127 citations counted in INSPIRE as of 26 Jul 2017


\bibitem{Gu:1999ks}
  Y.~F.~Gu and X.~H.~Li,
  %``Ratio of psi(2S) to J / psi hadronic decay rates and rho pi puzzle,''
  Phys.\ Rev.\ D {\bf 63}, 114019 (2001)
  %%doi:10.1103/PhysRevD.63.114019
  %%[hep-ph/9910406].
  %%CITATION = doi:10.1103/PhysRevD.63.114019;%%
  %37 citations counted in INSPIRE as of 26 Jul 2017

\bibitem{Brambilla:2010cs}
  N.~Brambilla {\it et al.},
  %``Heavy quarkonium: progress, puzzles, and opportunities,''
  Eur.\ Phys.\ J.\ C {\bf 71}, 1534 (2011)
  %%doi:10.1140/epjc/s10052-010-1534-9
  %%[arXiv:1010.5827 [hep-ph]].
  %%CITATION = doi:10.1140/epjc/s10052-010-1534-9;%%
  %1146 citations counted in INSPIRE as of 26 Jul 2017

\bibitem{Wang:2012mf}
  Q.~Wang, G.~Li and Q.~Zhao,
  %``Open charm effects in the explanation of the long-standing '$\rho\pi$ puzzle',''
  Phys.\ Rev.\ D {\bf 85}, 074015 (2012)
  %%doi:10.1103/PhysRevD.85.074015
  %%[arXiv:1201.1681 [hep-ph]].
  %%CITATION = doi:10.1103/PhysRevD.85.074015;%%
  %17 citations counted in INSPIRE as of 26 Jul 2017

\bibitem{bes3}
  M.~Ablikim {\it et al.} (BESIII Collaboration),
  Nucl.\ Instrum.\ Methods Phys. Rev. Sect.\ A {\bf 614}, 345 (2010).


\bibitem{KKMC}
  S.~Jadach, B.~F.~L.~ Ward and Z.~Was, Comput. Phys. Commun. {\bf 130},
  260 (2000); Phys. Rev. D {\bf 63}, 113009 (2001).

\bibitem{Lange:2001uf}  D.~J.~Lange,  %``The EvtGen particle decay simulation package,''
  Nucl.\ Instrum.\ Methods Phys. Rev. Sect. \ A {\bf 462}, 152 (2001);
  %%doi:10.1016/S0168-9002(01)00089-4
  %%CITATION = doi:10.1016/S0168-9002(01)00089-4;%%
  %1723 citations counted in INSPIRE as of 18 Jul 2017
\bibitem{EvtGen}  R.~G.~Ping {\it et al.}, Chin. Phys. C {\bf 32}, 599 (2008).

\bibitem{Chen:2000tv}
  J.~C.~Chen, G.~S.~Huang, X.~R.~Qi, D.~H.~Zhang and Y.~S.~Zhu,
  Phys.\ Rev.\  D {\bf 62}, 034003 (2000).
  %%CITATION = PHRVA,D62,034003;%%

\bibitem{geant4} S.~Agostinelli {\it et al.}
({\sc geant4} Collaboration), Nucl.\ Instrum.\ Methods Phys. Rev. Sect. \ A {\bf 506},
250 (2003).

\bibitem{Deng:2006} Z.~Y.~Deng {\it et al.}, High Energy Physics and Nuclear Physics {\bf 30}, 371 (2006).

\bibitem{root} http://root.cern.ch

\bibitem{tmva} A.~Hoecker, P.~Speckmayer, J.~Stelzer,
        J.~Therhaag, E.~von Toerne, and H.~Voss,
        ``TMVA: Toolkit for Multivariate Data Analysis,''
        PoS ACAT 040 (2007) [arxiv:0703039 [physics]].

\bibitem{bdt} J.~R.~Quinlan, in Proceedings of the Thirteenth National Conference
on Artificial Intelligence, Portland, Oregon, 1996.

\bibitem{Berger:2010my}
  N.~Berger, K.~Zhu, Z.~-A.~Liu, D.~-P.~Jin, H.~Xu, W.~-X.~Gong, K.~Wang and G.~-F.~Cao,
  %``Trigger efficiencies at BES III,''
  Chin.\ Phys.\ C {\bf 34}, 1779 (2010)
  %%[arXiv:1011.2825 [hep-ex]].
  %%CITATION = ARXIV:1011.2825;%%


\bibitem{Pacetti:2015iqa}
  S.~Pacetti, R.~Baldini Ferroli and E.~Tomasi-Gustafsson,
  %``Proton electromagnetic form factors: Basic notions, present achievements and future perspectives,''
  Phys.\ Rep.\  {\bf 550-551}, 1 (2015).
  % doi:10.1016/j.physrep.2014.09.005
  %%CITATION = doi:10.1016/j.physrep.2014.09.005;%%

\bibitem{Faldt:2017kgy}
  G.~F\"{a}ldt and A.~Kupsc,
  %``Hadronic structure functions in the $e^+ e^- \rightarrow \bar{\Lambda} \Lambda$ reaction,''
  Phys.\ Lett.\ B {\bf 772}, 16 (2017)
  % doi:10.1016/j.physletb.2017.06.011
  % [arXiv:1702.07288 [hep-ph]].
  %%CITATION = doi:10.1016/j.physletb.2017.06.011;%%
  %4 citations counted in INSPIRE as of 17 Jul 2018

\end{thebibliography}
\end{document}